\documentclass[sigconf,final]{acmart}

\usepackage{graphicx}
\usepackage{xcolor}
\usepackage{subfigure}
\usepackage{multirow}
\usepackage{enumitem}
\usepackage{pifont}

\usepackage{mathtools}

\usepackage[norelsize,linesnumbered,ruled,lined,boxed,commentsnumbered]{algorithm2e}

\acmConference[DAC'24]{Proceedings of the 59th ACM/IEEE Design Automation Conference}{July 10–14}{San Francisco, CA, USA} 
\acmYear{2024}
\copyrightyear{2024}

\acmPrice{15.00}

\begin{document}

\title{TITAN: A Distributed Large-Scale Trapped-Ion NISQ Computer}

\author{
\centering
\begin{tabular}{ccccccc}
Cheng Chu$^\dag$ & Zhenxiao Fu$^\dag$ & Yilun Xu$^\ddag$ & Gang Huang$^\ddag$ & Hausi Muller$^*$ & Fan Chen$^\dag$ & Lei Jiang$^\dag$
\end{tabular}
\begin{tabular}{ccc}
$^\dag$Indiana University & $^\ddag$Lawrence Berkeley National Laboratory & $^*$ University of Victoria\\
$^\dag$\textit{\{chu6, zhfu, fc7, jiang60\}@iu.edu}  & $^\ddag$\textit{\{yilunxu, ghuang\}@lbl.gov}  & $^*$\textit{hausi@uvic.ca}
\end{tabular}
}

\renewcommand{\authors}{Cheng Chu, Zhenxiao Fu, Yilun Xu, Gang Huang, Hausi Muller, Fan Chen, and Lei Jiang}
\renewcommand{\shortauthors}{C. Chu et al.}

\begin{abstract}
Trapped-Ion (TI) technology offers potential breakthroughs for Noisy Intermediate Scale Quantum (NISQ) computing. TI qubits offer extended coherence times and high gate fidelity, making them appealing for large-scale NISQ computers. Constructing such computers demands a distributed architecture connecting Quantum Charge Coupled Devices (QCCDs) via quantum matter-links and photonic switches. However, current distributed TI NISQ computers face hardware and system challenges. Entangling qubits across a photonic switch introduces significant latency, while existing compilers generate suboptimal mappings due to their unawareness of the interconnection topology. In this paper, we introduce TITAN, a large-scale distributed TI NISQ computer, which employs an innovative photonic interconnection design to reduce entanglement latency and an advanced partitioning and mapping algorithm to optimize matter-link communications. Our evaluations show that TITAN greatly enhances quantum application performance by $56.6\%$ and fidelity by $19.7\%$ compared to existing systems.
\end{abstract}

\ccsdesc[500]{Computer systems organization~Distributed architectures}
\ccsdesc[500]{Hardware~Quantum technologies}

\keywords{Trapped Ions, NISQ, Distributed Quantum Computers}

\maketitle

\section{Introduction}
\label{s:intro}

TI technology emerges as a promising avenue for the construction of large-scale NISQ computers, potentially unlocking quantum advantages~\cite{Daley:Nature2022}. TI qubits present several distinct advantages: longer coherence times, up to one hour~\cite{Wang:NC2021} compared to conventional superconducting qubits; higher gate fidelity, with 2-qubit gates achieving 99.92\% fidelity~\cite{Gaebler:PRL2016}; dense qubit connectivity for efficient 2-qubit gate implementation~\cite{Kielpinski:Nature2002}; and modular, scalable QCCDs~\cite{Mose:ARXIV2023}, exemplified by the latest Quantinuum's 32-qubit QCCD~\cite{Mose:ARXIV2023}.

To construct large-scale TI NISQ computers, a distributed architecture is essential. Integrating numerous qubits within a single QCCD device significantly degrades TI gate fidelity~\cite{Murali:ISCA2020}. While quantum matter-links~\cite{Akhtar:NC2023} can connect QCCDs into a TI module, excessive QCCD integration leads to cross-talk between QCCDs~\cite{Monroe:Science2013} and increased control and cooling overhead~\cite{Castillo:ACC2023}. Therefore, large-scale TI NISQ computers connect distributed QCCD-based TI modules via a photonic switch~\cite{Monroe:PRA2014}.

However, state-of-the-art distributed TI NISQ computers face challenges from both hardware and system perspectives. On the hardware side, entangling two qubits across a photonic switch requires a significant latency, $\sim60\times$ longer than a 2-qubit gate~\cite{Monroe:PRA2014}, impacting quantum application performance. The entanglement process involves establishing optical connections, cooling, and $\sim10$ entanglement attempts, each lasting $500\mu s$~\cite{Stephenson:PRL2020}. While more switch ports can reduce the final step's latency, they prolong optical connection establishment time. Thus, simply adding ports does not reduce overall entanglement latency. On the system side, state-of-the-art compilers~\cite{Baker:CF2020,Wu:MICRO2022} designed for distributed quantum computers generate unoptimized qubit mappings, due to their lack of awareness regarding the interconnection. These compilers minimize inter-module communications by graph partitioning, but overlook the presence of quantum matter-links, as well as the specific locations of photonic ports within each TI module. Consequently, the mappings generated by these compilers inadvertently lead to frequent communications across quantum matter-links.

In this paper, we propose \textit{TITAN}, a large-scale distributed TI NISQ computer, where multiple QCCDs are interconnected as a TI module by quantum matter-links, and multiple distributed TI modules are interconnected using a photonic switch. Our contributions are summarized as follows.
\begin{itemize}[leftmargin=*, nosep, topsep=0pt, partopsep=0pt]
\item \textbf{Innovative Photonic Interconnection Design}. We introduce an innovative photonic interconnection design for TITAN to accelerate entanglements across a photonic switch. By increasing the number of ports within each QCCD-based TI module, TITAN enhances the concurrency of entanglement attempts, thus reducing the latency associated with entangling qubits across a photonic switch. Instead of using a single large, slow photonic switch, TITAN employs multiple smaller, faster photonic switches to connect TI modules. 

\item \textbf{Advanced Partitioning and Mapping Algorithm}. We present a novel partitioning and mapping algorithm to curtail communication overhead across quantum matter-links. Our algorithm minimizes inter-module communications and inter-QCCD communications within each TI module through hierarchical partitioning. Furthermore, the algorithm optimizes communication patterns across quantum matter-links by allocating the qubit partition characterized by the highest frequency of inter-module communications to the QCCD positioned nearest to the photonic port within a TI module. 

\item \textbf{Enhanced Performance and Fidelity}. We evaluated and compared TITAN against existing distributed TI NISQ computers with previous compiler support. Our assessments demonstrate that, compared to previous TI NISQ computers, TITAN improves the performance and fidelity of various quantum applications by $56.6\%$ and $19.7\%$, respectively.
\end{itemize}

\begin{figure}[t!]
\centering
\includegraphics[width=0.9\linewidth]{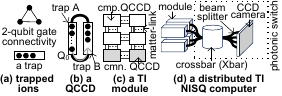}
\Description[short description]{long description}
\vspace{-0.1in}
\caption{The basics of TI technology, a trap, a QCCD, a module, and a distributed architecture.}
\label{f:q_circle_qccd}
\vspace{-0.3in}
\end{figure}

\vspace{-0.1in}
\section{Background and Motivation}
\label{s:back}

\subsection{Trapped-Ion Technology}

\textbf{Ion Trap and Gate}. In Trapped-Ion (TI) quantum systems, information is encoded within ions confined within an ion trap~\cite{Wang:NC2021}. Electrode segments at each end of the trap create a spatial confinement, while a radio-frequency electric field induces fluctuations, arranging ions into a linear chain, as shown in Figure~\ref{f:q_circle_qccd}(a). Quantum gates are realized through laser manipulation. Single-qubit gates involve interactions with specific ions, while 2-qubit gates require multiple lasers to excite both internal states and ion chain motion, enabling full qubit connectivity. The M{\o}lmer-S{\o}rensen (MS) gate the canonical 2-qubit gate, known for lower fidelity compared to single-qubit gates in TI systems~\cite{Gaebler:PRL2016}.

\textbf{Quantum Charge Coupled Device}. The fidelity of qubits is compromised when controlling and implementing quantum gates in a long ion chain~\cite{Murali:ISCA2020}, making a single-trap architecture unsuitable for scalability. To address this issue, a modular and scalable Quantum Charge Coupled Device (QCCD)~\cite{Mose:ARXIV2023} is introduced for constructing large-scale TI NISQ computers, as illustrated in Figure~\ref{f:q_circle_qccd}(b). The QCCD consists of two ion traps, each with a small number of ions, interconnected by conveyor belt regions~\cite{Mose:ARXIV2023}. To physically move a qubit ($Q_0$) from one trap ($A$) to the other ($B$), a split operation separates $Q_0$ from the ion chain in $A$, followed by a shuttling operation to move $Q_0$ to $B$ and a merge operation to combine it with the ion chain in $B$. Finally, a 2-qubit gate can occur between $Q_0$ and another qubit in $B$.

\textbf{TI Module}. A TI module comprises multiple QCCDs interconnected via quantum matter-links~\cite{Akhtar:NC2023}, as depicted in Figure~\ref{f:q_circle_qccd}(c). These quantum matter-links facilitate ion transfer between QCCDs through electric fields. At the QCCD boundary, an RF electrode aligns with the corresponding electrode of the adjacent QCCD. By applying translating potentials across the inter-QCCD gap, ions can be transported. The transportation of a qubit through a matter-link takes $0.4ms$ and attains a fidelity of $99.999993\%$~\cite{Akhtar:NC2023}. To minimize cross-talk, specific QCCDs are configured with more $^{171}$Yb$^+$ ions for computing (cmp.), while others emphasize $^{138}$Ba$^+$ ions, optimized for communication (cmn.) tasks~\cite{Brown:NQI2016}.

\vspace{-0.1in}
\subsection{Distributed TI NISQ Computer}

\textbf{Distributed TI NISQ Computer}. A large-scale TI NISQ computer is constructed by interconnecting multiple distributed TI modules using a photonic switch, as shown in Figure~\ref{f:q_circle_qccd}(d). A photonic switch~\cite{Monroe:Science2013} comprises components such as a MEMS optical crossbar (Xbar), beam splitters, and a CCD camera detector array. It enables the heralded and probabilistic distribution of entanglement between two TI modules. 

\begin{figure}[t!]
\centering
\includegraphics[width=\linewidth]{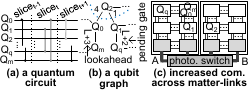}
\Description[short description]{long description}
\vspace{-0.3in}
\caption{Prior compilers on a distributed TI NISQ computer.}
\label{f:q_distribute_comp}
\vspace{-0.2in}
\end{figure}

\textbf{Entanglement via Photonic Switch}. Entangling qubits across a photonic switch~\cite{Stephenson:PRL2020} involves three phases: optical connection establishment, a cooling operation, and multiple entanglement attempts. The process begins with the ``Xbar operation'', where the photonic switch's Xbar component establishes an optical connection between incoming and outgoing ports. Subsequently, the photonic switch proceeds to the ``non-Xbar operations'', which include a cooling operation lasting $\sim100 \mu s$, followed by multiple entanglement attempts, each taking $500\mu s$~\cite{Stephenson:PRL2020}. These attempts are crucial for achieving a successful entanglement. Completing all non-Xbar operations for a single entanglement requires $\sim5.4ms$~\cite{Brown:NQI2016}, involving about 10 attempts. Notably, an entanglement incurs a significant latency, $\sim60\times$ longer than that of a TI 2-qubit gate, which adversely affects the performance of distributed TI NISQ computers.

\vspace{-0.1in}
\subsection{Compiler for Distributed NISQ Computer}

\textbf{Qubit Graph}. A qubit graph~\cite{Wu:MICRO2022,Baker:CF2020} plays a crucial role in optimizing mappings for quantum circuits in distributed NISQ computers. A circuit is divided into multiple time-slices, each containing a set of concurrently executed gates, as depicted in Figure~\ref{f:q_distribute_comp}(a). These time-slices are then abstracted into qubit graphs, with nodes representing data qubits and edges weighted to denote the number of 2-qubit gates, as highlighted in Figure~\ref{f:q_distribute_comp}(b).

\textbf{Lookahead Weight}. Relying solely on information from the current time-slice often results in suboptimal mappings. A \textit{lookahead} mechanism~\cite{Wu:MICRO2022,Baker:CF2020} is introduced to construct qubit graphs that span a more extended temporal range within the quantum circuit by enriching these qubit graphs with lookahead weights. The process of generating a qubit graph at time $t$ begins with the original qubit graph in time slice $t$ and assigns a super large weight (``L'') to edges connecting interacting qubits in the current time slice, guaranteeing that any mapping strategy will position these qubits within the same partition. For each qubit pair, the weight~\cite{Wu:MICRO2022,Baker:CF2020} of their edge is computed as follows:
\begin{equation}
w_t(Q_i, Q_j) = \sum_{t<m<T} I(m, Q_i, Q_j) \cdot D(m - t),
\end{equation}
where $D$ denotes an exponential decay function, i.e., $D(x)=2^{-x/\sigma}$, $I(m, Q_i, Q_j)$ is an indicator variable (equal to 1 if $Q_i$ and $Q_j$ interact in time slice $m$; and 0 otherwise), and $T$ represents the total number of time-slices in the circuit. This augmentation of qubit graphs with lookahead weights effectively takes into account the influence of upcoming time-slices, giving larger weights to interactions occurring sooner in the circuit.

\textbf{Graph Partitioning and Mapping}. Previous compilers~\cite{Wu:MICRO2022,Baker:CF2020}, designed for distributed quantum computing setups that encompass various NISQ computing units linked via a central hub, employ a recurring procedure involving the Kernighan-Lin algorithm. This process is geared towards the division of the qubit graph into several partitions of uniform size. Subsequently, each partition is indiscriminately assigned to one of the NISQ computing devices.

\vspace{-0.1in}
\subsection{Motivation}
Unfortunately, the performance and fidelity of state-of-the-art distributed TI NISQ computers are constrained by a combination of hardware- and system-level challenges. 

\begin{figure}[t!]
\centering
\subfigure[The Xbar operations in a larger Xbar.]{
   \includegraphics[width=1.5in]{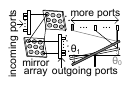}
	 \Description[q_xbar_hard]{long description}
   \label{f:q_xbar_hard}
}
\hspace{-0.1in}
\subfigure[The overall entanglement latency.]{
   \includegraphics[width=1.6in]{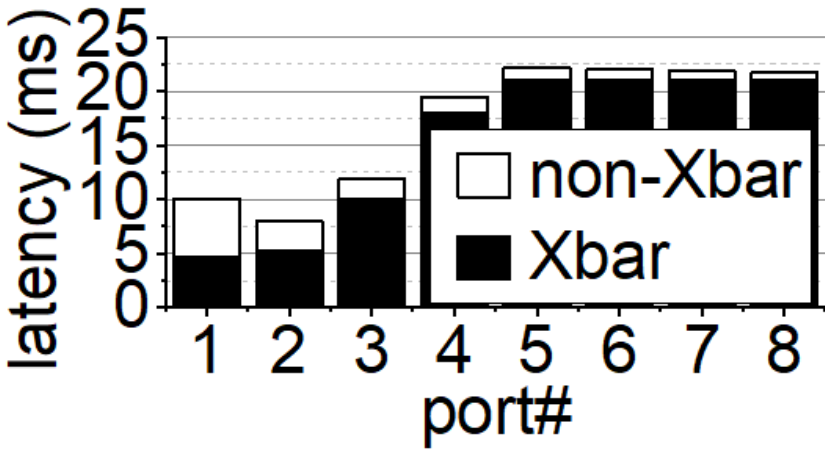}
	 \Description[q_port_num]{long description}
   \label{f:q_port_num}
}
\vspace{-0.2in}
\caption{Adding more ports for larger entanglement attempt concurrency in an entanglement.}
\label{f:q_switch_all}
\vspace{-0.3in}
\end{figure}

\textbf{Slow Photonic Switch}. A key impediment leading to significant entanglement latency is found in the final phase, i.e., the non-Xbar operations, which comprises multiple entanglement attempts, each consuming $500\mu s$~\cite{Stephenson:PRL2020}. Simply increasing the number of ports within the Xbar component of the photonic switch to enhance entanglement attempt concurrency does not effectively alleviate the overall entanglement latency. This is due to the fact that a larger Xbar, housing additional ports, substantially prolongs the duration required for establishing optical connections within the Xbar, i.e., the Xbar operation~\cite{Mellette:JLT2015}. As Figure~\ref{f:q_xbar_hard} illustrates, the Xbar necessitates a prolonged latency to rotate the mirror arrays by a larger angle ($\theta_1>\theta_0$) to accommodate an increased number of incoming and outgoing ports~\cite{Mellette:JLT2015}. While Figure~\ref{f:q_port_num} suggests that the latency of non-Xbar operations diminishes with an increasing number of Xbar ports, this advantage is offset by the protracted duration of the Xbar operation induced by the enlarged Xbar. 

\textbf{Compiler Limitations in Interconnection Handling}. Previous compilers~\cite{Wu:MICRO2022,Baker:CF2020}, designed for general distributed NISQ computers, while adept at minimizing communications between different TI modules, inadvertently amplify qubit movements across quantum matter-links, due to their lack of awareness regarding the interconnection of a distributed TI NISQ computer. As shown in Figure~\ref{f:q_distribute_comp}(b), these compilers effectively partition the qubit graph into two segments and subsequently map each segment to a TI module, achieving great reductions in inter-module communications. However, Figure~\ref{f:q_distribute_comp}(c) illustrates a potential pitfall. In some instances, these compilers may allocate two qubits, denoted as $Q_1$ and $Q_q$, recognized for their frequent inter-QCCD communications within module $A$, to two separate QCCDs that lack direct connectivity. This mapping choice results in an increased volume of ion movements across matter-links. The underlying cause of this issue lies in the absence of support for QCCD-level partitioning and mapping within each TI module by these previous compilers. Furthermore, another scenario arises where these compilers may allocate two qubits, $Q_1$ and $Q_2$, which are characterized by frequent inter-module communications, to two computing QCCDs positioned in separate TI modules without direct connections to their respective photonic ports. Consequently, this mapping strategy inevitably introduces additional communication loads across matter-links between a computing QCCD and a communication QCCD within each TI module.

\begin{figure}[t!]
\centering
\includegraphics[width=0.9\linewidth]{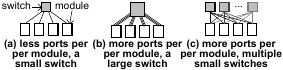}
\Description[short description]{long description}
\vspace{-0.15in}
\caption{The photonic interconnection in TITAN.}
\label{f:q_port_switch}
\vspace{-0.25in}
\end{figure}

\vspace{-0.1in}
\section{TITAN}
\label{s:titan}

In this paper, we present \textit{TITAN}, a rapid distributed TI NISQ computer that interconnects multiple QCCDs as a TI module using quantum matter-links, and further links multiple distributed TI modules through a photonic switch. TITAN has two innovative features: a novel photonic interconnection design and an advanced partitioning and mapping algorithm. Firstly, TITAN revolutionizes photonic interconnections to significantly expedite entanglement processes across the photonic switch. By augmenting the number of ports within each TI module, TITAN greatly improves the concurrency of entanglement attempts. This enhancement effectively reduces the latency associated with qubit entanglement through the photonic switch. Unlike conventional approaches employing single large and slow photonic switches, TITAN adopts a more efficient strategy with multiple smaller, faster photonic switches for interconnecting TI modules. Secondly, TITAN incorporates an innovative algorithm for partitioning and mapping. Our algorithm minimizes inter-module communications and inter-QCCD communications within each TI module through hierarchical partitioning. And it also optimizes communication patterns across matter-links by mapping the qubit partition with the highest frequency of inter-module communications to the communication QCCD situated nearest to the photonic port within a TI module. Our partitioning and mapping algorithm ensures a more optimized distribution of qubits, leading to a substantial reduction in communications that traverse quantum matter-links.

\begin{figure*}[t!]
\centering
\includegraphics[width=0.85\linewidth]{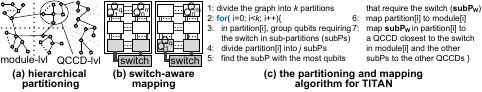}
\Description[short description]{long description}
\vspace{-0.15in}
\caption{The compiler design for TITAN comprising distributed TI modules interconnected by a photonic switch.}
\label{f:q_compile_set}
\vspace{-0.2in}
\end{figure*}

\vspace{-0.1in}
\subsection{Innovative Photonic Interconnection Design}

\textbf{More Ports in a TI Module}. TITAN leverages a photonic switch~\cite{Monroe:Science2013} to interconnect multiple distributed TI modules. To establish an entanglement, the photonic switch activates its optical Xbar, enabling optical connections between two TI modules. The time needed to configure these connections is referred to as the Xbar latency. Subsequently, the source TI module undergoes ion cooling ($\sim100 \mu s$) and executes around $\sim10$ entanglement attempts, each consuming $500\mu s$~\cite{Stephenson:PRL2020}, to successfully complete the entanglement with the destination TI module across the photonic switch. This duration, encompassing cooling and multiple entanglement attempts, is known as the non-Xbar latency. In previous distributed TI NISQ computers~\cite{Monroe:Science2013}, each TI module was equipped with only a limited number of ports, and a small, high-speed photonic switch was utilized to interconnect them, as depicted in Figure~\ref{f:q_port_switch}(a). However, the restricted number of ports in a TI module severely curtails the concurrency of entanglement attempts in an entanglement, substantially extending the non-Xbar latency. As illustrated in Figure~\ref{f:q_port_switch}(b), we propose a significant augmentation of the port count within each TI module in TITAN to amplify entanglement attempt concurrency and, consequently, reduce the non-Xbar latency. Nevertheless, the integration of more ports into a TI module significantly enlarges the Xbar size of the photonic switch, consequently slowing down the Xbar latency of an entanglement. Therefore, the mere addition of ports to a TI module is insufficient to reduce the overall latency of an entanglement, as highlighted in Figure~\ref{f:q_port_num}.

\textbf{Multiple Compact Photonic Switches}. To diminish both Xbar and non-Xbar latencies within an entanglement, TITAN adopts multiple smaller photonic switches to connect its TI modules, as illustrated in Figure~\ref{f:q_port_switch}(c). Suppose a prior TI distributed NISQ computer comprises $n$ TI modules, each equipped with $m$ ports (where $n \geq 2$ and $m \geq 1$), the interconnection necessitates a large $nm\times nm$ photonic switch. TITAN, on the other hand, augments the number of ports within each TI module by a factor of $10\times$, thereby endowing each TI module with $10m$ ports. Instead of using a single $10nm \times 10nm$ photonic switch, TITAN adopts $10m$ individual $n \times n$ photonic switches to connect the TI modules. This approach ensures that each port of a TI module is connected to a distinct photonic switch, as it is impossible for the ports of the same TI module to communicate with each other. The $n \times n$ photonic switch, responsible for interconnecting the many-ported TI modules within TITAN, exhibits a more compact form, thereby rendering it swifter compared to the $nm\times nm$ photonic switch employed by the earlier distributed TI NISQ computer.

\vspace{-0.1in}
\subsection{Partitioning and Mapping}
The pseudocode of our partitioning and mapping algorithm is described in Figure~\ref{f:q_compile_set}(c).

\textbf{Hierarchical Partitioning}. Previous compilers for distributed quantum computers~\cite{Wu:MICRO2022,Baker:CF2020} employ the Kernighan-Lin algorithm to partition a qubit graph into multiple equally-sized partitions, each subsequently mapped to a dedicated NISQ computer. However, this approach cannot be straightforwardly applied to partition a qubit graph for a distributed TI NISQ computer, which comprises $k$ TI modules, each housing $j$ QCCDs. One naive approach might involve dividing the qubit graph into $k \cdot j$ partitions of equal size and mapping each to a designated QCCD. However, this naive solution would require invoking the Kernighan-Lin algorithm for $q \cdot C_{j\cdot k}^{2}$ times, where $q$ is the number of optimization iterations. The Kernighan-Lin algorithm has a time complexity of $\mathcal{O}(pn^2\log n)$~\cite{Baker:CF2020}, where $n$ is the number of qubits, and $p$ represents the number of optimization iterations. Consequently, the time complexity of this naive solution scales as $\mathcal{O}(q \cdot C_{j\cdot k}^{2} pn^2\log n)$. When $k$, $j$, and $n$ are sufficiently large, this approach becomes impractically slow. In contrast, as Figure~\ref{f:q_compile_set}(a) shows, we propose a \textit{hierarchical partitioning} technique to efficiently divide the qubit graph into $k\cdot j$ partitions. Similar to previous compilers, we initially partition the qubit graph into $k$ partitions of equal size using the Kernighan-Lin algorithm (Line 1 of Figure~\ref{f:q_compile_set}(c)). Each partition is then mapped to a specific TI module (Line 6). Subsequently, we further divide each partition into $j$ smaller sub-partitions of equal size (Line 4), with each sub-partition being assigned to a QCCD within the corresponding TI module (Line 7). The foundation of our hierarchical partitioning lies in the recognition that the latency of the photonic switch connecting TI modules exceeds that of the quantum matter-links interconnecting QCCDs within each TI module. This hierarchical partitioning approach maintains a time complexity of $\mathcal{O}((kpn^2\log n + jp(n/k)^2\log(n/k))$, much smaller than the na\"ive solution. The goal of our hierarchical partitioning is to group two logic qubits frequently interacting with each other in a sub-partition. For instance, unlike the inefficient partitions made by previous compilers and shown in Figure~\ref{f:q_distribute_comp}(c), as Figure~\ref{f:q_compile_set}(b) shows, our hierarchical partitioning can group $Q_1$ and $Q_q$ sharing a large weight in one sub-partition, and $Q_0$ and $Q_m$ joining a pending 2-qubit gate in another sub-partition to reduce the inter-QCCD communications within each TI module.

\textbf{Switch-aware Mapping}. Previous compilers tailored for distributed NISQ computers~\cite{Wu:MICRO2022,Baker:CF2020} have a uniform mapping approach where each partition is assigned to a distributed NISQ computer without considering any specific attributes. This approach is suitable when dealing with the mapping of large partitions of the qubit graph to TI modules. However, directly applying this uniform approach when mapping smaller sub-partitions to individual QCCDs results in suboptimal mappings. This is due to variations in the distances between each QCCD and the photonic ports within a TI module. To address this issue, we propose a \textit{switch-aware} mapping scheme that optimizes the placement of sub-partitions within a TI module to minimize ion movements. Our switch-aware mapping approach takes into account the unique characteristics of each TI module's interconnection setup. In our switch-aware mapping, after the module-level separation achieved through hierarchical partitioning, instead of randomly initializing all sub-partitions, we first group qubits that communicate frequently with other partitions into several sub-partitions (Line 3 of Figure~\ref{f:q_compile_set}(c)), while the remaining sub-partitions are initialized randomly. After the hierarchical partitioning, our switch-aware mapping strategy identifies the sub-partition with the highest frequency of communication with other partitions within each partition (Line 5). It then assigns this sub-partition to the communication QCCDs that are closest to the photonic ports within the TI module (Line 7). The other sub-partitions are mapped to the remaining QCCDs in the TI module. As illustrated in Figure~\ref{f:q_compile_set}(b), for instance, the sub-partition comprising qubits $Q_0$ and $Q_m$ exhibits the most frequent communication with other partitions. Our switch-aware mapping approach, rather than assigning this sub-partition to a QCCD in the first row of the TI module, places it in one of the communication QCCD closest to the photonic ports, which, in this case, is the last row of the TI module. 


\begin{table}[t!]
\centering
\caption{The design overhead comparison.}
\vspace{-0.1in}
\setlength{\tabcolsep}{3pt}
\footnotesize
\begin{tabular}{|c||l|}
\hline
Scheme                     & Description \\\hline\hline
\multirow{6}{*}{baseline}  & 4 TI modules are interconnected by a $256\times256$ photonic switch. A\\
                           & module has 64 photonic ports and consists of 6 QCCDs, each having up\\
                           & to 32 qubits. 8 matter-links connect two neighboring QCCDs in a TI\\
                           & module. For an entanglement across the switch, Xbar - $5.23ms$ \&\\
                           & non-Xbar - $2.75ms$. 512 physical qubits: 256 data qubits \& 256 for com.\\\hline
													
\multirow{3}{*}{TITAN}     & TI module and QCCD configurations are the same as baseline. TITAN\\
                           & employs 8 $32\times32$ switches to interconnect 4 TI modules. For an entang-\\
                           & lement across a switch, Xbar - $1.1ms$ \& non-Xbar - $0.765ms$.\\\hline
\end{tabular}
\label{f:quantum_comp_all}
\vspace{-0.15in}
\end{table}

\begin{table}[t!]
\centering
\caption{The simulated benchmarks}
\vspace{-0.1in}
\setlength{\tabcolsep}{3pt}
\footnotesize
\begin{tabular}{|c||c|c|c|c|}
\hline
benchmark                & logic qubit     & 2-qubit gate \#     \\ \hline\hline
Adder (ADD)              & 256             & 2033                \\
Bernstein–Vazirani (BV)  & 256             & 255                \\
QAOA (QAO)               & 256             & 1020                \\
Quantum Primacy (PRI)    & 256             & 192               \\
Random (RAN)             & 256             & 2705                \\
Hamiltonian (HAM)        & 256             & 510                \\\hline
\end{tabular}
\label{t:quantum_simulated_benchmark}
\vspace{-0.2in}
\end{table}

\section{Experimental Methodology}
\label{s:em}

\textbf{Baseline Configuration}. The hardware configuration of our baseline is shown in Table~\ref{f:quantum_comp_all}. We have adopted the most recent racetrack QCCD design~\cite{Mose:ARXIV2023}, accommodating up to 32 qubits. Six QCCDs are interconnected as a TI module, with four QCCDs dedicated to computing tasks and the remaining two tailored for communication functions. Each QCCD employs 8 quantum matter-links~\cite{Akhtar:NC2023} to establish connections with neighboring QCCDs. The transportation of ions through a matter-link takes $0.4ms$, attaining a fidelity of $99.999993\%$~\cite{Akhtar:NC2023}. Within each TI module, there are 64 photonic ports. Based on Figure~\ref{f:q_port_num}, an entanglement simultaneously performs two entanglement attempts to obtain the minimal overall latency. A module supports two concurrent entanglements. Moreover, we assume three distillation iterations for each entanglement, elevating its fidelity from an initial 94\%~\cite{Stephenson:PRL2020} to a target of 99.3\%, incurring an 8-qubit/port overhead. These four TI modules are interconnected via a $256\times256$ photonic switch. For entanglement operations across the switch, the Xbar latency is $5.23ms$~\cite{Mellette:JLT2015}, while non-Xbar operations require $2.75ms$~\cite{Stephenson:PRL2020}. Our baseline is composed of 512 physical qubits distributed across six QCCDs, with 256 designated for data storage and processing and the remaining 256 allocated to entanglements and their distillations. To compile quantum applications within our baseline, we employ a state-of-the-art compiler~\cite{Baker:CF2020} specifically designed for general distributed NISQ computers. However, it is important to note that this compiler primarily supports module-level partitioning and mapping. Consequently, our baseline directly maps qubits within a partition to the four QCCDs within a TI module, based on their natural order.

\textbf{Design Overhead}. The design overhead is presented in Table~\ref{f:quantum_comp_all}. TITAN shares the same hardware configurations as our baseline, except the follows. Although a module also owns 64 photonic ports, it supports an entanglement by 8 concurrent entanglement attempts. TITAN employs 8 $32\times32$ photonic switches to interconnect its four TI modules. Compared to the $256\times256$ switch, the 8 $32\times32$ switches of TITAN reduce the size of mirror arrays by 87.5\%. For an entanglement across a switch, the Xbar latency is $1.1ms$~\cite{Mellette:JLT2015}, and the non-Xbar requires $0.765ms$. Compared to the compiler~\cite{Baker:CF2020} of our baseline, our TITAN's compiler has to perform QCCD-level partitioning and mapping having a time complexity of $\mathcal{O}((jp(n/k)^2\log(n/k))$, where $n$ is the number of qubits, $p$ represents the number of optimization iterations, $j$ is the number of QCCDs in a TI module, and $k$ is the number of TI modules.

\textbf{Quantum Applications}. Our study focuses on a range of quantum applications detailed in Table~\ref{t:quantum_simulated_benchmark}. In the context of our baseline, we specifically consider applications characterized by 256 data qubits and varying numbers of 2-qubit MS gates, spanning from 192 to 2.7K. \textit{Adder} (ADD)~\cite{Murali:ISCA2020} is frequently used in QFT and quantum phase estimation. \textit{Bernstein–Vazirani} (BV)~\cite{Wright:NC2019} determines an unknown bit string within a black-box function. \textit{QAOA} (QAO)~\cite{Moll:QST2018} addresses combinatorial optimization problems across diverse domains. \textit{Quantum Primacy} (PRI)~\cite{arute:2019nature} generates random circuits to showcase quantum advantage. \textit{Random} (RAN)~\cite{Huang:NJP2023} assembles quantum gates randomly, forming circuits for quantum machine learning. \textit{Hamiltonian} (HAM)~\cite{oftelie:PRB2020} generates circuits for simulating 1D Transverse Field Ising Models.

\begin{table}[t!]
\centering
\caption{The timing and fidelity models.}
\vspace{-0.1in}
\setlength{\tabcolsep}{3pt}
\footnotesize
\begin{tabular}{|c|c|c||c|c|c|}
\hline
operation    & time ($\mu s$)              & infidelity  & operation         & time ($\mu s$)               & infidelity  \\ \hline
1-qubit gate & 5~\cite{Gutierrez:PRA2019}  & 3e-5        & 2-qubit gate      & 100~\cite{Murali:ISCA2020}   & 8e-4        \\
merge/split  & 380~\cite{Gutierrez:PRA2019}& -           & 1-step shuttling  & 5~\cite{Murali:ISCA2020}     & 1e-5        \\
X-Junction   & 100~\cite{Blakestad:PRL2009}& 1e-4        & measurement       & 400~\cite{Gutierrez:PRA2019} & 9e-5        \\
matter-link  & 400~\cite{Akhtar:NC2023}    & 7e-8        & photonic switch   & 5760                         & 7e-3        \\\hline
\end{tabular}
\label{t:quantum_time_model}
\vspace{-0.2in}
\end{table}

\textbf{Timing \& Fidelity}. Our timing and fidelity models (Table~\ref{t:quantum_time_model}) rely on latency and fidelity values from~\cite{Murali:ISCA2020} for 2-qubit gates and shuttling, and from~\cite{Gutierrez:PRA2019} for 1-qubit gates, merge/split, and measurement. A typical merge/split operation takes $80\mu s$\cite{Gutierrez:PRA2019} but may reduce gate fidelity due to ion chain heating. To mitigate this, cooling operations during merging/splitting require $\sim300\mu s$\cite{feng:PRL2020}. The X-Junction design~\cite{Blakestad:PRL2009} and matter-link design~\cite{Akhtar:NC2023} are applied. Entanglement latency across a photonic switch is calculated in Table~\ref{f:quantum_comp_all}. 

\textbf{Simulation}. We modified and extended a state-of-the-art simulator~\cite{Murali:ISCA2020} designed for a single QCCD to model multi-QCCD TI modules and distributed TI NISQ computers. Our modified simulator utilizes IBM's Qiskit framework for circuit processing and benchmark implementation.

\vspace{-0.1in}
\section{Evaluation}
\label{s:eval}

\begin{figure}[t!]
\centering
\includegraphics[width=\linewidth]{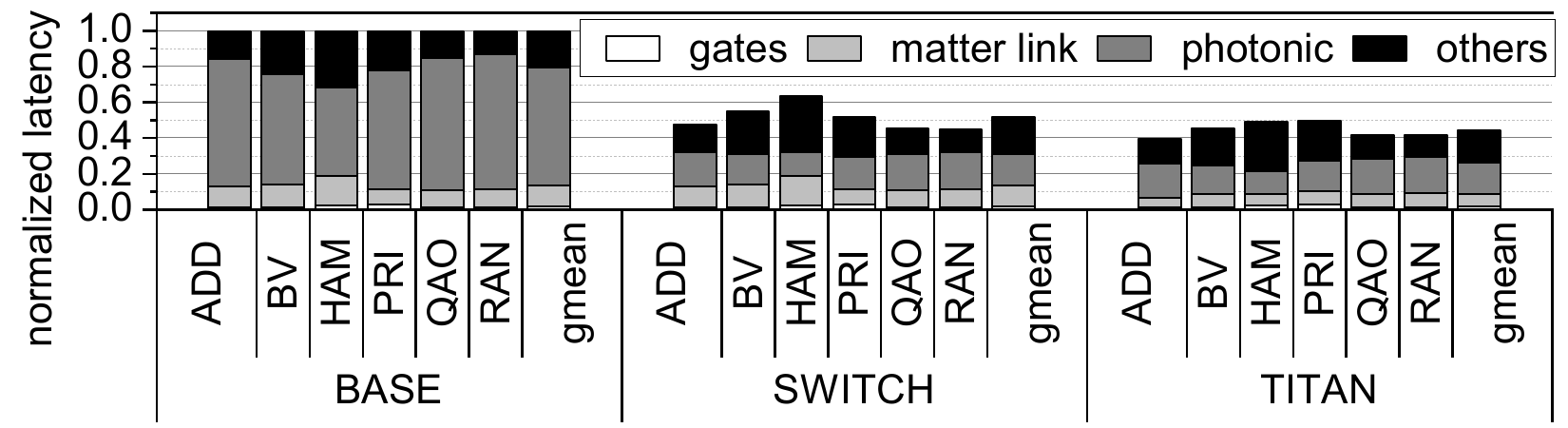}
\vspace{-0.25in}
\caption{The quantum application performance of TITAN.}
\label{f:performance_comparison}
\vspace{-0.2in}
\end{figure}

\textbf{Performance}. The performance of various quantum applications achieved by TITAN is shown in Figure~\ref{f:performance_comparison}. In our baseline (BASE), entanglements through the photonic switch constitutes an average of $66.2\%$ of the total application latency, due to the long latency of the large switch. On average, our new photonic interconnection design (SWITCH) featured by small, low-latency Xbars yields a significant performance improvement of $48.6\%$. Through both hierarchical partitioning and switch-aware mapping of TITAN, on average, the application performance is furhter increased by $13.1\%$ over SWITCH, due to the reduction of matter-link usage between QCCDs in each module. Meanwhile, the diminished ion movements across matter-links also decrease the frequency of various TI operations, including merge/split, shuttle, and X-junction. Overall, the amalgamation of both hardware and system techniques in TITAN results in a $56.6\%$ increase in quantum application performance compared to BASE.

\begin{figure}[h!]
\vspace{-0.1in}
\centering
\includegraphics[width=\linewidth]{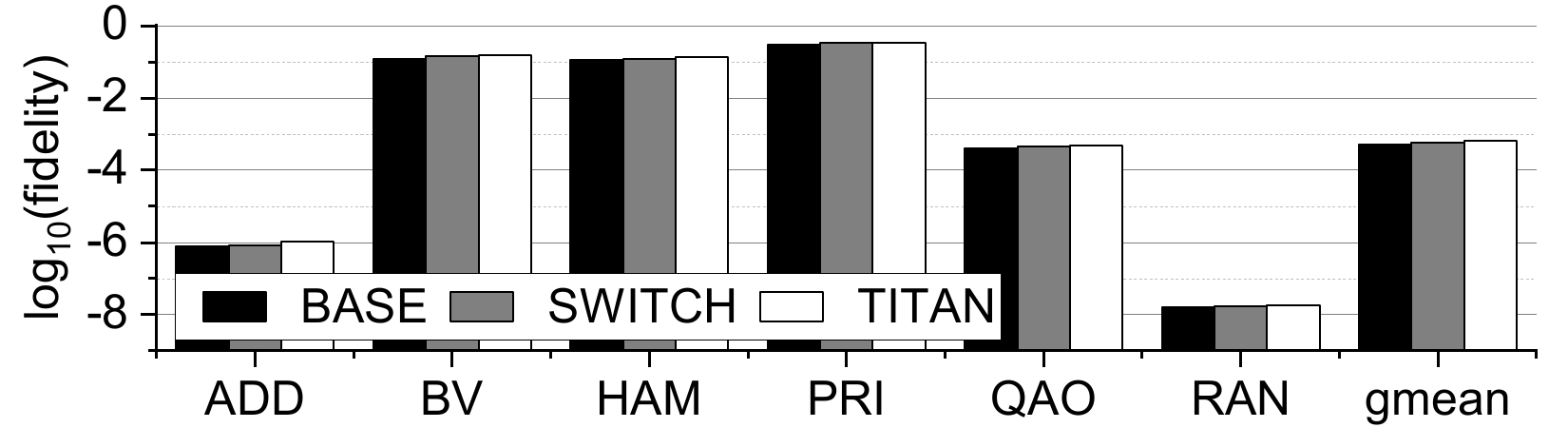}
\vspace{-0.2in}
\caption{The quantum application fidelity of TITAN.}
\label{f:fidelity_comparison}
\vspace{-0.1in}
\end{figure}

\textbf{Fidelity}. The fidelity of quantum applications of TITAN is exhibited in Figure~\ref{f:fidelity_comparison}. On average, SWITCH exhibits a $14.2\%$ improvement in fidelity compared to BASE, due to its shorter application latency. The evolution of the state $|\psi \rangle$ of a quantum system can be described as $i\hbar\frac{d|\psi \rangle}{dt}=H|\psi \rangle$, where $H$ is the Hamiltonian determining the evolution, $\hbar$ is the reduced Planck constant, and $i$ is the imaginary unit. A reduced application latency leads to less decoherence and relaxation of quantum states, thereby improving fidelity even when the operation type and number remain unchanged. TITAN further reduces the TI operation count and noises in quantum circuits by hierarchical partitioning and switch-aware mapping, resulting in a $4.8\%$ fidelity enhancement across all benchmarks. Overall, TITAN achieves a fidelity improvement of $19.7\%$.

\begin{figure}[t!]
\centering
\begin{minipage}[b]{0.495\linewidth}
\centering
\includegraphics[width=1.5in]{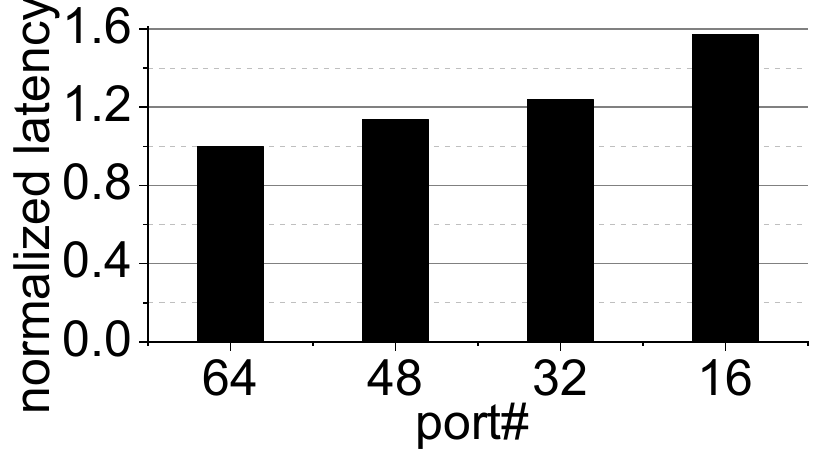}\vspace{-2pt}
\vspace{-0.12in}
\caption{Study on Port \#}\label{f:partition_performance}
\end{minipage}\vspace{-8pt}
\hspace{-0.05in}
\begin{minipage}[b]{0.495\linewidth}
\centering
\includegraphics[width=1.5in]{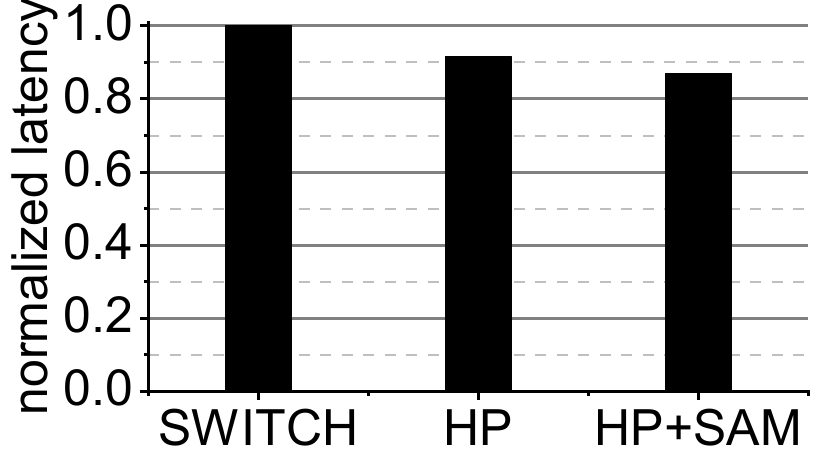}\vspace{-2pt}    
\vspace{-0.12in}
\caption{Partition \& Mapping}\label{f:ports_performance}
\end{minipage}
\end{figure}

\textbf{Partition \& Mapping}. Figure~\ref{f:partition_performance} illustrates the performance improvement achieved by hierarchical partitioning and switch-aware mapping of TITAN, with all results normalized to SWITCH. On average, hierarchical partitioning (HP) demonstrates a 8.5$\%$ performance improvement compared to SWITCH, since HP groups qubits frequently communicating with each other in a QCCD. The combination of HP and switch-aware mapping (SAM) achieves a 5.2$\%$ performance improvement over HP, since SAM maps qubits requiring entanglements to the communication QCCDs close to photonic ports. Overall, two schemes together decreases the number of ion movements across matter-links, leading to a 13.1$\%$ performance enhancement.

\textbf{Sensitivity Study on Port \#}. Figure~\ref{f:ports_performance} depicts the performance variations across different port numbers in the photonic interconnection design of TITAN. All results are normalized to BASE with 64 ports. As the port number decreases, there is a corresponding increase in the latency of each benchmark. In comparison to BASE, the latencies for 48 ports, 32 ports, and 16 ports exhibit average increments of 13.46$\%$, 23.67$\%$, and 57.01$\%$, respectively. This rise in latency is predominantly because of the reduced number of ports in each TI module, diminishing the concurrency of entanglement attempts and consequently leading to prolonged latency in every photonic entanglement.

\section{Conclusion}
\label{s:con}

In this paper, we present TITAN, a large-scale distributed TI NISQ computer featured by an innovative photonic interconnection design and an advanced partitioning and mapping algorithm. Our results show TITAN greatly enhances quantum application performance by $56.6\%$ and fidelity by $19.7\%$ compared to existing systems.

\begin{acks}
This work was supported in part by NSF CCF-2105972, and NSF CAREER AWARD CNS-2143120. We thank the IonQ Research Credits Program for their support.
\end{acks}

\bibliographystyle{acma}
\bibliography{quantum}
\end{document}